\documentclass[11pt]{article}
\pdfoutput=1

\usepackage{aas_macros,amsmath,amssymb,cite,esint,graphicx,multirow,subcaption}
\usepackage[margin=.8in,letterpaper]{geometry}
\usepackage[colorlinks=true]{hyperref}
\usepackage[affil-it]{authblk}

\numberwithin{equation}{section}
\let\originalleft\left
\let\originalright\right
\renewcommand{\left}{\mathopen{}\mathclose\bgroup\originalleft}
\renewcommand{\right}{\aftergroup\egroup\originalright}
\mathcode`\*="8000
{\catcode`\*=\active\gdef*{\mathclose{}\,\mathopen{}}}

\newcommand{\ab}[1]{\left|#1\right|}

\newcommand{\br}[1]{\left[#1\right]}

\newcommand{\pa}[1]{\left(#1\right)}

\newcommand{\ed}{\mathop{}\!\mathrm{d}}
\newcommand{\pd}{\mathop{}\!\partial}
\renewcommand{\O}[1]{\mathcal{O}\pa{#1}}

\begin{document}

\title{\Huge{Critical Emission from a High-Spin Black Hole}}
\date{}
\author[1,2]{Alexandru Lupsasca\thanks{lupsasca@fas.harvard.edu}}
\author[3]{Achilleas P. Porfyriadis\thanks{app@physics.ucsb.edu}}
\author[1]{Yichen Shi\thanks{yshi@g.harvard.edu}}
\affil[1]{\small Center for the Fundamental Laws of Nature, Harvard University, Cambridge, MA 02138, USA}
\affil[2]{\small Society of Fellows, Harvard University, Cambridge, MA 02138, USA}
\affil[3]{\small Department of Physics, UCSB, Santa Barbara, CA 93106, USA}

\maketitle

\begin{abstract}
    We consider a rapidly spinning black hole surrounded by an equatorial, geometrically thin, slowly accreting disk that is stationary and axisymmetric. We analytically compute the broadening of electromagnetic line emissions from the innermost part of the disk, which resides in the near-horizon region. The result is independent of the disk's surface emissivity and therefore universal. This is an example of critical behavior in astronomy that is potentially observable by current or future telescopes.
\end{abstract}

\thispagestyle{empty}\vfill\pagebreak

\setcounter{page}{1}

\tableofcontents

\section{Introduction}

With the historic detection of gravitational waves from black hole mergers \cite{Abbott2016a,Abbott2016b,Abbott2017a,Abbott2017b,Abbott2017c}, LIGO has ushered in a new era of data in observational black hole astrophysics. Now another compelling experiment, the Event Horizon Telescope (EHT), promises to soon deliver the first up-close picture of the black hole at the center of our Galaxy \cite{Broderick2006,Doeleman2008a,Doeleman2008b,Doeleman2009,Doeleman2012,Johnson2015}. These and other proposed experiments, such as ATHENA, SKA, and LISA, are bringing observational black hole astrophysics to a qualitatively new level of precision. This exciting development poses a pressing question to theorists: What exactly will these experiments observe?

While the predictions for gravitational wave signals are based on firm theoretical grounds, the observational signatures of black holes in various telescopes depend sensitively on their surroundings and can be influenced by the structure of their magnetosphere, as well as myriad other elements possibly present in their environment (such as a corona, jets, $etc$.), each of which carries its own degrees of freedom. Therefore, when making predictions for observations by telescopes, the common strategy is to produce many templates covering as much of this complex parameter space as possible. In contrast, one could profitably undertake a different approach and investigate the inverse problem: Is there a region in parameter space that would lead to such a distinctive signature that its observation could leave no doubt as to the nature of the source?

A distinctive corner in the parameter space is occupied by rapidly spinning black holes that saturate the Kerr bound for a black hole's angular momentum with respect to its mass, $J\leq M^2$. In this regime, an enhancement of symmetry in the immediate vicinity of the black hole's horizon \cite{Bardeen1999,Guica2009} allows for an analytic study of a variety of potentially observable phenomena \cite{Porfyriadis2014a,Porfyriadis2014b,Li2015,Lupsasca2014,Bai2014,Lupsasca2015,Hadar2015,Gralla2015,Compere2015,Gralla2016a,Gralla2016b,Porfyriadis2017,Burko2016,Wei2017,Hadar2017,Compere2017a,Gralla2017,Gralla2017b,Compere2017b}, which often exhibit a striking universality. A characteristic example, which is also relevant for the computation in this paper, is the observation \cite{Bardeen1973, Porfyriadis2017,Gralla2017} that all of the light emitted from the near-vicinity of a rapidly spinning black hole is constrained to appear on the so-called ``NHEKline": a vertical line segment on the edge of the shadow of every high-spin black hole. The authors of Ref.~\cite{Gralla2017} mapped the primary image on the NHEKline corresponding to an isotropically emitting point source that is orbiting near the horizon. In this paper, we analytically compute the broadening of electromagnetic line emissions observed along the NHEKline and originating from the innermost part of a radiant accretion disk around a high-spin black hole. Remarkably, we find that the result is independent of the disk's surface emissivity and therefore universal. Our prediction is summarized in Sec.~\ref{sec:Summary}.

Before turning to a technical summary of our results, it is worthwhile to present the high-energy theory viewpoint from which we approach this and related projects. Much of the quest for a fundamental theory of quantum gravity has revolved around the AdS/CFT correspondence. This relation studies gravity in asymptotically anti-de Sitter spacetimes whose geometry describes a ``gravitating box." One might therefore think that such systems are not available for study in nature. However, surprisingly, they are: The region of spacetime near the horizon of a high-spin black hole is an AdS-like region in our sky. Furthermore, it is now experimentally accessible. More precisely, a maximally spinning Kerr black hole is a critical point of the Kerr family for which the region of spacetime in the vicinity of the event horizon becomes an AdS-like vacuum solution of Einstein's equations in its own right \cite{Bardeen1999}. This so-called Near-Horizon Extremal Kerr (NHEK) geometry exhibits an enhanced $\mathsf{SL}(2,\mathbb{R})$ isometry group. Moreover, properties of diffeomorphisms in general relativity imply that this geometrically realized global conformal symmetry extends to an even larger infinite-dimensional local conformal symmetry. This fact lies at the root of the Kerr/CFT correspondence \cite{Guica2009}. Over the past few years, the action of these symmetries has rendered feasible a large number of analytical computations of astrophysically relevant processes in NHEK that could not otherwise be performed, sometimes not even numerically\cite{Porfyriadis2014a,Porfyriadis2014b,Li2015,Lupsasca2014,Bai2014,Lupsasca2015,Hadar2015,Gralla2015,Compere2015,Gralla2016a,Gralla2016b,Porfyriadis2017,Burko2016,Wei2017,Hadar2017,Compere2017a,Gralla2017,Gralla2017b,Compere2017b}. Because extremal black holes are examples of critical conformal fixed points in astronomy, they often display universal behavior. This paper predicts a new and striking example of universal critical behavior, which opens the tantalizing possibility of detecting a ``smoking gun" for conformal symmetry in the sky.

\subsection{Summary of results}
\label{sec:Summary}

In this paper, we consider a rotating Kerr black hole of mass $M$ and angular momentum $J=aM$. We specialize to a high-spin black hole that is close to saturating the Kerr bound $J\le M^2$. We surround the black hole with a geometrically thin, stationary, axisymmetric, equatorial disk of slowly accreting matter, and we assume that every particle in the disk emits monochromatic light isotropically in the form of photons that follow null geodesics. As reviewed in Sec.~\ref{sec:Setup}, the method of geometric optics may then be used to obtain the flux observed at infinity as a function of the photons' redshift, $F_o(g)$. In Sec.~\ref{sec:CriticalPoint}, we use this method to treat photon trajectories originating from the NHEK region in extreme Kerr and we obtain an analytic formula for $F_o(g)$. Up to a proportionality constant, the result is independent of the disk model and therefore universal. Section~\ref{sec:NearExtreme} generalizes the result to the case of a near-extreme Kerr. In Sec.~\ref{sec:SymmetricModel}, we propose a model for a radiant disk that respects the symmetries of NHEK. Due to the logarithmic divergence of the disk's proper length at extremality, this model implies a logarithmically divergent overall proportionality constant in $F_o(g)$.

We now state the main result of the paper. Suppose the observer's screen has Cartesian coordinates $(\alpha,\beta)$ and is located at a dimensionless coordinate distance $r_o\gg 1$ from the black hole, at a polar angle $\theta_o$ with respect to the hole's rotation axis. The observer receives nonvanishing flux $F_o(g)$ provided $\theta_o\in(\theta_c,\pi-\theta_c)$, with $\theta_c=\arccos\sqrt{2\sqrt{3}-3}$, and $g\in\left(1/\sqrt3,\sqrt3\right)$. The flux is given by
\begin{align}
	\label{eq:MainResult}
	F_o\propto\frac{1}{M^3r_o^2\sin{\theta_o}}\frac{g\pa{g-1/\sqrt{3}}\pa{g+\sqrt{3}}}{\sqrt{4g^2\pa{3+\cos^2{\theta_o}-4\cot^2{\theta_o}}-3\pa{g-1/\sqrt{3}}\pa{5g+\sqrt{3}}}}.
\end{align}
This result holds independently of the details of the disk (which only enter through the proportionality constant) and is therefore universal. The result \eqref{eq:MainResult} is directly relevant for the profile of $\mathrm{FeK}\alpha$ line emissions which have been extensively analyzed in \cite{Laor1991,Beckwith2004,Dovciak2004,Brenneman2006,Dauser2010} (see also, $e.g.$, the reviews \cite{Fabian2000,Reynolds2003,Brenneman2013}). As such, this result could pertain to spectral observations by experiments such as XMM-Newton, Suzaku, and NuSTAR, which have revealed emissions from high-spin black holes such as MCG-6-30-15 ($a\gtrsim0.98M$), NGC 1365 ($a\gtrsim0.97M$), and NGC 4051 ($a\gtrsim0.99M$) \cite{Tanaka1995,Pounds2004,Risaliti2013,Fabian2016}. Furthermore, we find that the flux $F_o$ emerges on the NHEKline, defined on the observer's screen by \cite{Gralla2017}
\begin{align}
    \label{eq:NHEKline}
    \alpha=-2M\csc{\theta_o},\qquad
    \ab{\beta}<\beta_\mathrm{max}=M\sqrt{3+\cos^2{\theta_o}-4\cot^2{\theta_o}},
\end{align}
at an elevation $\beta$ which is completely fixed by the redshift $g$ according to
\begin{align}
    \beta=\pm M\sqrt{3+\cos^2{\theta_o}-4\cot^2{\theta_o}-\frac{3}{4g^2}\pa{g-1/\sqrt{3}}\pa{5g+\sqrt{3}}}.
\end{align}
It follows that, for a suitable source line, a high-resolution telescope such as EHT might be able to observe the critical flux \ref{eq:MainResult} as a characteristic brightness profile along the NHEKline.

\begin{figure}[!ht]
	\centering
	\includegraphics[width=.37\textwidth]{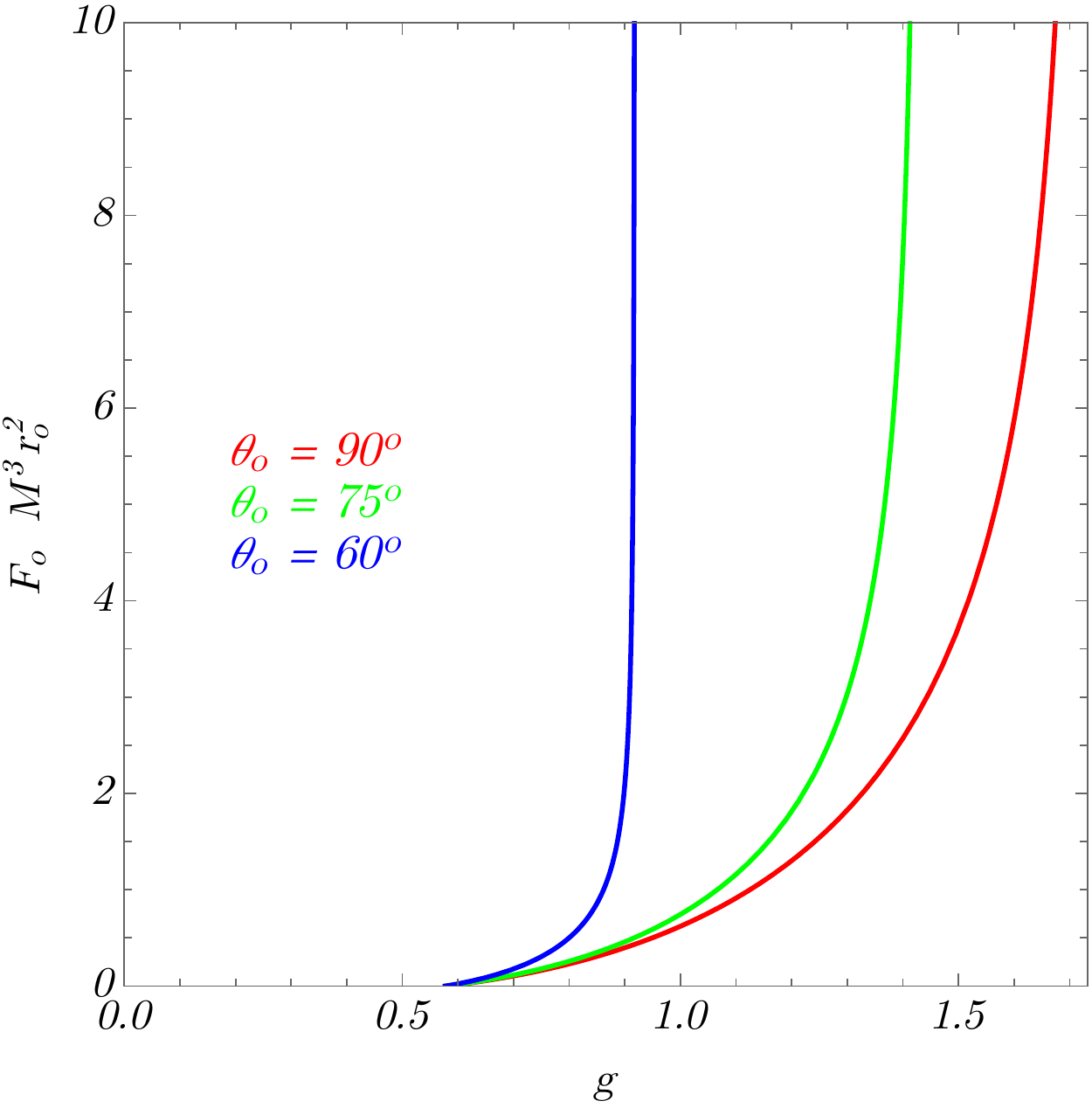}
	\caption{The profile of line emissions from the near-horizon region of a (near-)extremal black hole [Eq.~\eqref{eq:MainResult}]. This is our main result: an analytic expression for the flux $F_o$ measured by an observer at dimensionless radius $r_o$ and polar angle $\theta_o$ from the hole as a function of the redshift $g$. The result is independent of the disk's surface emissivity and therefore universal. This is an example of critical behavior in astronomy. Note that all the divergences in these plots are integrable. They arise from emissions that are aimed directly at the observer, in that they appear at the center of the NHEKline shown in Fig.~\eqref{fig:NHEKline}.}
	\label{fig:RedshiftEmissionProfile}
\end{figure}

\begin{figure}[!ht]
	\centering
	\begin{subfigure}[t]{.37\textwidth}
		\centering
		\includegraphics[width=\textwidth]{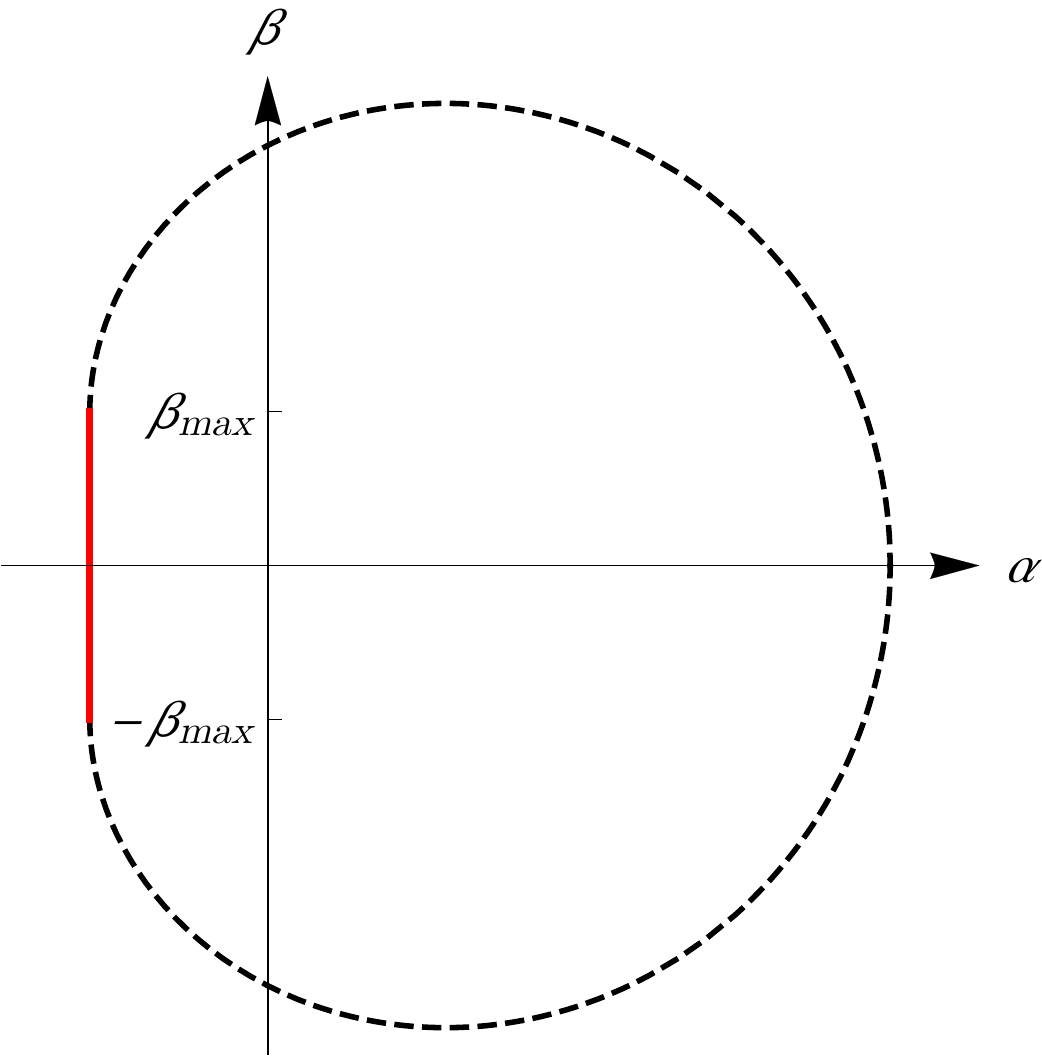}
		\caption{NHEKline of a high-spin black hole}
		\label{fig:NHEKline}
	\end{subfigure}
	\qquad
	\begin{subfigure}[t]{.37\textwidth}
		\centering
		\includegraphics[width=\textwidth]{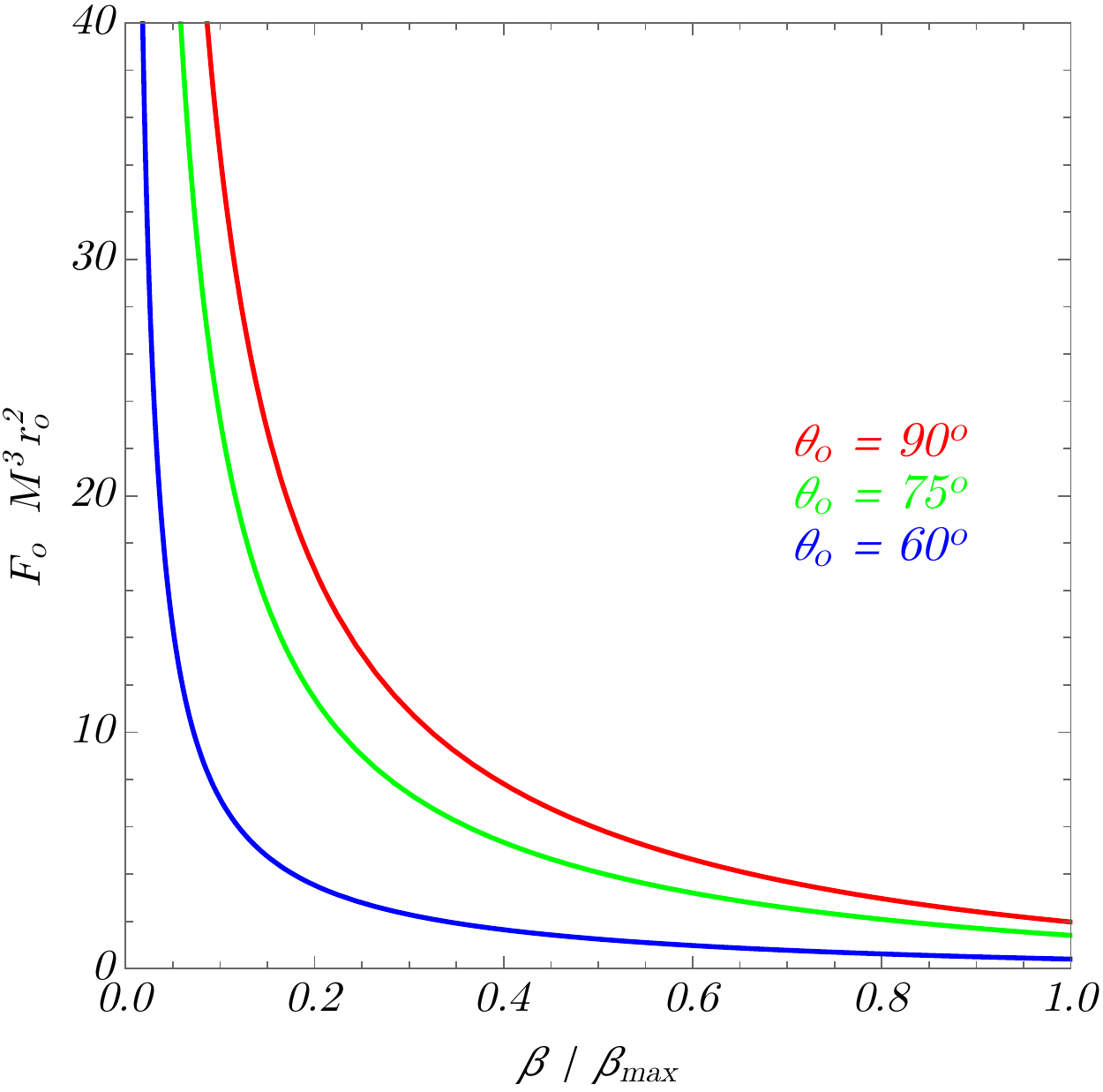}
		\caption{Universal critical emission profile}
		\label{fig:ScreenEmissionProfile}
	\end{subfigure}
	\caption{In the high-spin regime, the shadow that a black hole casts on a distant observer's screen develops a vertical edge: the so-called ``NHEKline" [Eq.~\eqref{eq:NHEKline}], depicted in red in the left panel. All electromagnetic emissions from the near-horizon region, including those from the innermost part of its accretion disk, are constrained to emerge on this vertical line segment. The line emissions from the near-horizon region computed in this paper may be thought of as brightness profiles along the NHEKline and we plot them as such on the right panel. An observation of the NHEKline with such a brightness profile would provide a ``smoking gun" of conformal symmetry in the sky. The divergences are likely due to the caustics discussed in Ref.~\cite{Gralla2017}, which should be regulated by diffraction effects beyond our geometric optics approximation.}
	\label{fig:Results}
\end{figure}

\section{Electromagnetic line emissions from a black hole accretion disk}
\label{sec:Setup}

The profile of electromagnetic line emissions from a disk of matter accreting onto a black hole is commonly computed via the geometric optics methods developed by Bardeen and Cunningham \cite{Cunningham1972,Cunningham1973,Bardeen1973,Cunningham1975}. In this section, we review how this is done for the case of emissions originating from a slowly accreting equatorial disk that is geometrically thin.

Astrophysically realistic black holes are described by the Kerr family of metrics, parameterized by their mass $M$ and angular momentum $J=aM$. In Boyer-Lindquist coordinates, the Kerr line element is
\begin{align}
\label{eq:Kerr}
    ds^2=-\frac{\Delta}{\Sigma}\pa{\ed t-a\sin^2{\theta}\ed\phi}^2+\frac{\Sigma}{\Delta}\ed \hat{r}^2+\Sigma\ed\theta^2+
\frac{\sin^2{\theta}}{\Sigma}\br{\pa{\hat{r}^2+a^2}\ed\phi-a\ed t}^2,
\end{align}
where
\begin{align}
    \Delta(\hat{r})=\hat{r}^2-2M\hat{r}+a^2,\qquad
    \Sigma(\hat{r},\theta)=\hat{r}^2+a^2\cos^2{\theta}.
\end{align}
A particle orbiting on a prograde, circular, equatorial geodesic at radius $\hat{r}=\hat{r}_s$ has four-velocity \cite{Bardeen1972}
\begin{align}
\label{eq:EquatorialGeodesics}
    u_s=u_s^t\pa{\pd_t+\Omega_s\pd_\phi},\qquad
    u_s^t=\frac{\hat{r}_s^{3/2}+aM^{1/2}}{\sqrt{\hat{r}_s^3-3M\hat{r}_s^2+2aM^{1/2}\hat{r}_s^{3/2}}},\qquad
    \Omega_s=\frac{M^{1/2}}{\hat{r}_s^{3/2}+aM^{1/2}}.
\end{align}
Here and hereafter, the subscript $s$ stands for ``source." Such an orbit is stable as long as
\begin{align}
    \hat{r}_s\ge\hat{r}_\mathrm{ISCO}=M\pa{3+Z_2-\sqrt{\pa{3-Z_1}\pa{3+Z_1+2Z_2}}},
\end{align}
where $\hat{r}_\mathrm{ISCO}$ denotes the radius of the Innermost Stable Circular Orbit (ISCO), with
\begin{align}
    Z_1&=1+\pa{1-a_\star^2}^{1/3}\br{\pa{1+a_\star}^{1/3}+\pa{1-a_\star}^{1/3}},\qquad
    Z_2=\pa{3a_\star^2+Z_1^2}^{1/2},\qquad
    a_\star=\frac{a}{M}.
\end{align}

In this paper, we consider a Kerr black hole \eqref{eq:Kerr} surrounded by a thin accretion disk consisting of particles falling along the equatorial geodesics described by the four-velocity \eqref{eq:EquatorialGeodesics}. Their actual four-velocity may also have a vertical component for motion in and out of the equator, as well as a radial component providing the disk with a nonzero accretion rate. However, we will assume these components to be small relative to the angular velocity, so that we may treat the particles' orbits as circular. In the region outside the ISCO, this assumption is valid for slowly accreting disks ($e.g.$, it has been carefully established in the context of the Novikov-Thorne model \cite{Novikov1973,Page1974,Penna2012}).

The particles in the disk can emit radiation that flows along null geodesics to reach a distant observer at radius $r_o$ and polar angle $\theta_o$. Here and hereafter, the subscript $o$ stands for ``observer." By the reflection symmetry of the problem, we may assume without loss of generality that the observer lies in the northern hemisphere, $\theta_o\in\pa{0,\pi/2}$.\footnote{In this paper, we ignore the measure-zero, degenerate cases of a precisely ``face-on" ($\theta_o=0$) or precisely ``edge-on" ($\theta_o=\pi/2$) observer.} We also assume that the disk is stationary and axisymmetric, in which case it suffices to consider the null geodesic motion in the $(r,\theta)$ plane only.

Let $p$ denote the four-momentum of the null geodesic corresponding to a photon trajectory connecting a source point to the observer. For such geodesics, the energy $E=-p_t$ may be scaled out of the geodesic equation, whose solutions may therefore be labeled by
\begin{align}
    \label{eq:ImpactParameters}
    \hat{\lambda}=\frac{L}{E},\qquad
    \hat{q}=\frac{\sqrt{Q}}{E},
\end{align}
where $L=p_\phi$ denotes the component of angular momentum parallel to the axis of symmetry and $Q=p_\theta^2-\cos^2{\theta}(a^2p_t^2-p_\phi^2\csc^2{\theta})$ is the Carter constant.\footnote{Note that $\hat{q}$ is manifestly real for photons emitted from the equatorial plane.} The null geodesic equation in the $(r,\theta)$ plane is given by
\begin{align}
    \label{eq:KerrGeodesicEquation}
    \fint_{\hat{r}_s}^{\hat{r}_o}\frac{\ed\hat{r}}{\pm\sqrt{\hat{R}(\hat{r})}}=\fint_{\theta_s}^{\theta_o}\frac{\ed\theta}{\pm\sqrt{\hat{\Theta}(\theta)}},
\end{align} 
where 
\begin{subequations}
	\begin{align}
	\hat{R}(\hat{r})&=\big(\hat{r}^2+a^2-a\hat{\lambda}\big)^2-\Delta\big[\hat{q}^2+(a-\hat{\lambda})^2\big],\\
	\hat{\Theta}(\theta)&=\hat{q}^2+a^2\cos^2{\theta}-\hat{\lambda}^2\cot^2{\theta}.
	\end{align}
\end{subequations}
Here, the slash notation $\fint$ is meant to indicate that these integrals are line integrals along a trajectory connecting the source and observer, with the signs chosen so that the integrals grow secularly. Since the emitted energy of the photon is $E_s=-p\cdot u_s$ and the energy at the distant observer is the conserved quantity $E_o=E=-p_t$, the redshift factor is \cite{Cunningham1973}
\begin{align}
    \label{eq:Redshift}
    g=\frac{E_o}{E_s}
    =\frac{\sqrt{\hat{r}_s^3-3M\hat{r}_s^2+2aM^{1/2}\hat{r}_s^{3/2}}}{\hat{r}_s^{3/2}+M^{1/2}(a-\hat{\lambda})}.
\end{align}
This relation may be used to determine the conserved quantity $\hat{\lambda}$ in terms of $(\hat{r}_s,g)$. Then, in principle, the geodesic equation \eqref{eq:KerrGeodesicEquation} may be used to determine $\hat{q}$ also in terms of $(\hat{r}_s,g)$.

At the observer, the standard procedure involves defining impact parameters $(\alpha,\beta)$ in terms of $(\hat{\lambda},\hat{q})$ for every geodesic hitting the observer's screen \cite{Cunningham1972,Cunningham1973,Bardeen1973}:
\begin{align}
    \label{eq:AlphaBeta}
    \alpha=-\frac{\hat{\lambda}}{\sin{\theta_o}},\qquad
    \beta=\pm\sqrt{\hat{\Theta}(\theta_o)}.
\end{align}
A bundle of nearby geodesics that reach the screen from the disk then subtends a solid angle given by
\begin{align}
    \label{eq:SolidAngle}
    \ed\Omega=\frac{1}{\hat{r}_o^2}\ed\alpha\ed\beta
    =\frac{1}{\hat{r}_o^2}\ab{\frac{\pd(\alpha,\beta)}{\pd(\hat{\lambda},\hat{q})}}\ed\hat{\lambda}\ed\hat{q}
    =\frac{1}{\hat{r}_o^2}\ab{\frac{\pd(\alpha,\beta)}{\pd(\hat{\lambda},\hat{q})}}\ab{\frac{\pd(\hat{\lambda},\hat{q})}{\pd(\hat{r}_s,g)}}\ed\hat{r}_s\ed g.
\end{align}
The first Jacobian is straightforward to compute from Eq.~\eqref{eq:AlphaBeta}:
\begin{align}
    \label{eq:ObserverJacobian}
    \ab{\frac{\pd(\alpha,\beta)}{\pd(\hat{\lambda},\hat{q})}}=\frac{\hat{q}}{\sin{\theta_o}\ab{\beta}}.
\end{align}
On the other hand, the second Jacobian can typically only be computed numerically because no analytic expression relating $\hat{r}_s$, $g$, and $\hat{q}$ has been derived in the most general setting considered here.

Finally, the specific flux carried to the observer by the bundle of photons is given by
\begin{align}
    \label{eq:DifferentialFlux}
    \ed F_o=I_o\ed\Omega=g^3I_s\ed\Omega,
\end{align}
where Liouville's theorem on the invariance of the phase space density of photons has been used to relate the observed specific intensity $I_o$ to the emitted one $I_s$. We take the disk's specific intensity to be monochromatic at energy $E_\star$ ($e.g.$, $E_{\mathrm{FeK}\alpha}=6.38$keV), and isotropic with surface emissivity $\mathcal{E}(\hat{r}_s)$:
\begin{align}
    I_s=\mathcal{E}(\hat{r}_s)\delta\pa{E_s-E_\star}
    =g\mathcal{E}(\hat{r}_s)\delta\pa{E_o-gE_\star}.
\end{align}
Plugging this expression into Eq.~\eqref{eq:DifferentialFlux} and using Eqs.~\eqref{eq:SolidAngle}--\eqref{eq:ObserverJacobian}, the $\ed g$ integral corresponds to trivially setting $g=E_o/E_\star$, so we arrive at
\begin{align}
\label{eq:DiskFlux}
    F_o=\frac{g^4}{\hat{r}_o^2\sin{\theta_o}}\int\frac{\hat{q}}{\ab{\beta}}\ab{\frac{\pd(\hat{\lambda},\hat{q})}{\pd\pa{\hat{r}_s,g}}}\mathcal{E}(\hat{r}_s)\ed\hat{r}_s,
\end{align}
where we have absorbed a factor of $E_\star$ into $\mathcal{E}(r_s)$. Here, it is understood that the integral is to be evaluated over the radial extent of the accretion disk, typically starting from the ISCO. In the most general setting discussed so far, the result of the integral \eqref{eq:DiskFlux} will depend, via the surface emissivity $\mathcal{E}(\hat{r}_s)$, on the particular disk model ($e.g.$, Novikov-Thorne) that one chooses to employ for describing the accretion of matter into the black hole.\footnote{In practice, a (broken) power law $\mathcal{E}(r_s)\propto r_s^{-p}$ is often implemented when fitting data, with the power(s) chosen to best fit the data (see $e.g.$, Ref.~\cite{Brenneman2006}).} 

In the next sections, we will consider the regime where the black hole has (near-)maximal spin and the emissions come from the innermost portion of the accretion disk lying near the ISCO. In this regime, we will find an analytic expression for the Jacobian ${\pd(\hat{\lambda},\hat{q})}/{\pd(\hat{r}_s,g)}$ which will enable us to compute $F_o$ analytically. Moreover, we will find that, up to an overall (possibly infinite) constant, the answer is entirely independent of the disk's surface emissivity and therefore universal.

\section{Critical behavior of the maximally spinning extreme Kerr}
\label{sec:CriticalPoint}

In this section, we specialize to the critical point of the Kerr family of metrics: the $J=M^2$ extreme Kerr. Small deviations from extremality are considered in the next section. 

It is convenient to define shifted dimensionless radial coordinate and parameters \cite{Porfyriadis2017,Gralla2017}
\begin{align}
    r=\frac{\hat{r}-M}{M},\qquad
    \lambda=1-\frac{\hat{\lambda}}{2M},\qquad
    q^2=3-\frac{\hat{q}^2}{M^2},
\end{align}
in terms of which the geodesic equation \eqref{eq:KerrGeodesicEquation} becomes
\begin{align}
    \fint_{r_s}^{r_o}\frac{\ed r}{\sqrt{R(r)}}=\fint_{\theta_s}^{\theta_o}\frac{\ed\theta}{\sqrt{\Theta(\theta)}},
\end{align}
with
\begin{subequations}
\begin{align}
	R(r)&=r^4+4r^3+\pa{q^2+8\lambda-4\lambda^2}r^2+8\lambda r+4\lambda^2,\\
	\Theta(\theta)&=3-q^2+\cos^2{\theta}-4(1-\lambda)^2\cot^2{\theta}.
\end{align}    
\end{subequations}
When the source point $r_s$ is near the horizon ($r_s\ll1$) and the observation point $r_o$ is far in the asymptotically flat region of the spacetime ($r_o\gg1$), an analytical solution for the radial integral was recently found in Ref.~\cite{Porfyriadis2017}. In the same regime, an analytical expression for the polar integral was also recently obtained in Ref.~\cite{Gralla2017}. The analysis in Refs.~\cite{Porfyriadis2017,Gralla2017} shows that all electromagnetic signals from the near-horizon region $r_s\ll1$ appear on the observer's screen on the NHEKline with coordinates
\begin{align}
\label{eq:NHEKlineCoordinates}
    \alpha=-2M\csc{\theta_o},\qquad
    \beta=\pm M\sqrt{3-q^2+\cos^2{\theta_o}-4\cot^2{\theta_o}}.
\end{align}
Moreover, a key observation that may be derived semi-analytically from the results in Ref.~\cite{Gralla2017} is that for a fixed redshift $g$, the dominant contribution to the flux measured from sources at $r_s$ is achieved by photons emitted with conserved quantities $(\lambda,q)$ such that $r_s$ is a near-region radial turning point for the photon's geodesic. The near-region radial turning point is given by \cite{Porfyriadis2017,Gralla2017}
\begin{align}
\label{eq:TurningPoint}
    r_s=-\frac{2\lambda}{q^2}\pa{2+\sqrt{4-q^2}}.
\end{align}
On the other hand, for $r_s\ll1$, we obtain from Eq.~\eqref{eq:Redshift}
\begin{align}
    \label{eq:lambda(rs,g)}
    \lambda=-\frac{3r_s}{4g}\pa{g-1/\sqrt{3}}.
\end{align}
The parameter ranges are $\lambda<0$, $g\in(1/\sqrt{3},\sqrt{3})$, and $q\in(0,\sqrt{3+\cos^2{\theta_o}-4\cot^2{\theta_o}})$. Note that Eqs.~\eqref{eq:TurningPoint}--\eqref{eq:lambda(rs,g)} imply that $q$ depends only on $g$:
\begin{align}
    \label{eq:q(rs,g)}
    q=\frac{\sqrt{3}}{2g}\sqrt{\pa{g-1/\sqrt{3}}\pa{5g+\sqrt{3}}}.
\end{align}
Given Eq.~\eqref{eq:NHEKlineCoordinates}, this implies that $\beta$ also depends only on $g$ according to
\begin{align}
    \label{eq:beta(g)}
    \beta=\pm M\sqrt{3+\cos^2{\theta_o}-4\cot^2{\theta_o}-\frac{3}{4g^2}\pa{g-1/\sqrt{3}}\pa{5g+\sqrt{3}}},
\end{align}
meaning that the flux at different points on the observer's NHEKline is dominated by photons of different energy. From Eqs.~\eqref{eq:lambda(rs,g)}--\eqref{eq:q(rs,g)}, we may readily compute the Jacobian
\begin{align}
    \frac{\pd(\lambda,q)}{\pd(r_s,g)}=-\frac{3\sqrt{3}}{16qg^4}\pa{g-1/\sqrt{3}}\pa{g+\sqrt{3}}.
\end{align}
Using this in Eq.~\eqref{eq:DiskFlux}, we then find
\begin{align}
    \label{eq:FoResult}
    F_o=\frac{3\sqrt{3}\pa{g-1/\sqrt{3}}\pa{g+\sqrt{3}}}{8r_o^2\sin{\theta_o}\ab{\beta}}\int\mathcal{E}(r_s)\ed r_s.
\end{align}
This remarkable equation is the main result of the paper. Together with Eq.~\eqref{eq:beta(g)}, it gives an explicit analytic formula for the observed flux $F_o$ as a function of the redshift $g$ [Eq.~\eqref{eq:MainResult}], or equivalently, as a function of the elevation $\beta$ on the NHEKline. We plot these functions in Figs.~\ref{fig:RedshiftEmissionProfile} and \ref{fig:ScreenEmissionProfile}, respectively. The result is independent, up to an overall constant, of the particular disk model. The latter's role is merely to supply an emissivity function $\mathcal{E}$, whose integral fixes the overall scale of $F_o$. As we will see in Sec.~\ref{sec:SymmetricModel}, this overall scale might in fact be diverging at extremality.

It is worth emphasizing that the factorization of Eq.~\eqref{eq:DiskFlux} into the form of Eq.~\eqref{eq:FoResult} is consistent with expectations from conformal symmetry. Indeed, the $\mathsf{SL}(2,\mathbb{R})$ global conformal symmetry includes dilations and this implies that there are no special radii in the near-horizon part of the disk. As a result, one would expect that in the end the radial integration in Eq.~\eqref{eq:DiskFlux} trivializes as manifested in Eq.~\eqref{eq:FoResult}.

\section{Near-critical behavior of the near-extreme Kerr}\label{sec:NearExtreme}

In this section, we consider a near-extremal black hole with a small deviation from extremality measured by a parameter $\epsilon\ll1$ such that
\begin{align}
    a=M\sqrt{1-\epsilon^3}.
\end{align}
This choice of parametrization places the ISCO at an $\O{\epsilon}$ coordinate distance from the horizon: 
\begin{align}
    r_\mathrm{ISCO}=2^{1/3}\epsilon+\O{\epsilon^2}.
\end{align}

As has been previously observed in Ref.~\cite{Gralla2015} for gravitational wave fluxes from extreme-mass-ratio inspirals, and in Ref.~\cite{Gralla2017} for electromagnetic wave fluxes from an orbiting hot spot, the near-extreme result is often simply related to the extremal one by a natural identification of parameters. This is believed to be a manifestation of the action of the infinite-dimensional conformal group, which can relate extremal to near-extremal physics. Here the relevant identification of parameters is
\begin{align}
    r\to\epsilon r,\qquad
    \lambda\to\epsilon\lambda.
\end{align}
We have verified that with this identification, all the equations of the previous section concerning extreme Kerr are valid to leading order in $\epsilon$ for the near-extreme case as well. In particular, Eq.~\eqref{eq:FoResult} gives the leading-order observed flux from the portion of the accretion disk that is located at an $\O{\epsilon}$ coordinate distance from the horizon of a near-extreme Kerr black hole, whose deviation from extremality is given by $\epsilon$. Moreover, the flux from any other portion of the accretion disk that is at an $\O{\epsilon^p}$ coordinate distance from the horizon, with $0<p<3/2$, may be similarly obtained via the identification $r\to\epsilon^pr$, $\lambda\to\epsilon^p\lambda$.

\section{Symmetric model for a radiant disk: A conjecture}
\label{sec:SymmetricModel}

In the previous sections, we have seen that the profile of electromagnetic emissions from the innermost parts of an accretion disk surrounding a (near-)extremal Kerr is independent of the disk model, up to an overall constant that is given by the disk's integrated emissivity function $\mathcal{E}$. In this section, we motivate and propose a symmetric model for a homogeneous radiant disk. This symmetric disk model implies that the overall constant that multiplies the emission profile \eqref{eq:MainResult} diverges logarithmically as the black hole approaches extremality.

The source particle number current for an equatorial, stationary, axisymmetric disk that terminates at the ISCO takes the form
\begin{align}
    \label{eq:SourceCurrent}
	\mathcal{J}_s=\rho(\hat{r}_s)H(\hat{r}-\hat{r}_\mathrm{ISCO})\delta\pa{\theta-\pi/2}u_s,
\end{align}
where $H$ is the Heaviside step function. We assume that this source current, which is completely fixed by the radial density profile of emitters in the disk, $\rho(\hat{r}_s)$, is conserved: $\nabla_\mu\mathcal{J}_s^\mu=0$. We also assume that every particle emits isotropically, so that the local emissivity at the surface of the disk is $\mathcal{E}(\hat{r}_s)=\rho(\hat{r}_s).$

Near extremality, the emergent conformal symmetry of the throat geometry includes dilations. To preserve this symmetry, we assume that the disk has a uniform particle number density per unit proper radial length. This completely fixes the surface emissivity to
\begin{align}
	\label{eq:Emissivity}
	\mathcal{E}(\hat{r}_s)=\frac{1}{u_s^t\hat{r}_s\sqrt{\Delta(\hat{r}_s)}}
	=\frac{\sqrt{\hat{r}_s^3-3M\hat{r}_s^2+2aM^{1/2}\hat{r}_s^{3/2}}}{\hat{r}_s\pa{\hat{r}_s^{3/2}+aM^{1/2}}\sqrt{\hat{r}_s^2-2M\hat{r}_s+a^2}}.
\end{align}
Integrating this over radii $r_s$ that scale like $r_s\sim\epsilon^p$ produces a logarithmically divergent constant
\begin{align}
    \label{eq:DivergentFlux}
	\int\mathcal{E}(r_s)\ed r_s\sim\log\epsilon
\end{align}
that multiplies the profile \eqref{eq:MainResult}. This is due to the logarithmic divergence of the disk's proper length at extremality.

In the terminology of Ref.~\cite{Gralla2016a}, the choice of emissivity function \eqref{eq:Emissivity} makes the source current \eqref{eq:SourceCurrent} a vector field of weight $H=0$ on Kerr. As a result, its leading piece in a near-horizon expansion is completely determined by the symmetries of NHEK. Any other choice of disk model that has a well-behaved near-horizon limit will also have a source current of weight $H=0$ and therefore its leading behavior in the NHEK limit will match that of our symmetric model, with deviations arising only at subleading order (from near-horizon fields with larger conformal dimension). In particular, we expect the same logarithmic divergence \eqref{eq:DivergentFlux} in the flux received from any slowly accreting disk with a well-behaved near-horizon limit.

\section*{Acknowledgements}

The authors wish to thank Andrew Strominger for numerous helpful suggestions and comments on this and related projects. We also thank Geoffrey Comp\`ere, Samuel Gralla, and Roberto Oliveri for useful conversations. Y.~S. is grateful to Laura Brenneman, Thomas Dauser, and Michael Parker for providing data on various spinning black holes. A.~L. and Y.~S. are also grateful to Sheperd Doeleman and Michael D. Johnson for fruitful discussions at the Black Hole Initiative at Harvard University, which is supported by a grant from the John Templeton Foundation. This work was supported in part by the Sir Keith Murdoch Fellowship and NSF Grants No.~1205550 and No.~1504541.

\bibliography{Library}
\bibliographystyle{utphys}

\end{document}